\documentclass[%
reprint,
superscriptaddress,
 amsmath,amssymb,
 aps,
]{revtex4-2}

\usepackage{graphicx}
\usepackage{dcolumn}
\usepackage{bm}
\usepackage{xcolor}
\usepackage{hyperref}
\usepackage{enumitem,amssymb}
\usepackage{pifont}
\newlist{todolist}{itemize}{2}
\setlist[todolist]{label=$\square$}
\graphicspath{{Figures/}}
\usepackage{lineno}

\begin{document}
\preprint{}

\title{Helical jets driven by a ring of laser irradiation}

\author{Kian Orr}%
\email{kianorr@princeton.edu}
\affiliation{Department of Mechanical and Aerospace Engineering, Princeton University, Princeton, New Jersey 08544, USA}

\author{Brandon K. Russell}%
\affiliation{Department of Astrophysical Sciences, Princeton University, Princeton, New Jersey 08544, USA}

\author{Kirill Lezhnin}%
\affiliation{Princeton Plasma Physics Laboratory, Princeton University, 100 Stellarator Rd, Princeton, NJ, 08540, USA}

\author{Yang Zhang}%
\affiliation{Department of Astrophysical Sciences, Princeton University, Princeton, New Jersey 08544, USA}

\author{Geoffrey Pomraning}%
\affiliation{Department of Astrophysical Sciences, Princeton University, Princeton, New Jersey 08544, USA}

\author{Petros Tzeferacos}%
\affiliation{Department of Physics and Astronomy, University of Rochester, Rochester, NY 14627, United States of America}

\author{Hantao Ji}%
\affiliation{Department of Astrophysical Sciences, Princeton University, Princeton, New Jersey 08544, USA}
\affiliation{Princeton Plasma Physics Laboratory, Princeton University, 100 Stellarator Rd, Princeton, NJ, 08540, USA}

\author{Lan Gao}%
\affiliation{Princeton Plasma Physics Laboratory, Princeton University, 100 Stellarator Rd, Princeton, NJ, 08540, USA}

\date{\today}

\begin{abstract}

Plasma jets are formed in various astrophysical systems as plasma is rapidly ejected from a source, with a subset of these jets being magnetized and having a helical structure. Here, we demonstrate that helical jets may be formed using a ring of laser pulses that arrive on planar foils sequentially with increasing energy. The formation of the jets and their properties, including kinetic helicity, are studied through a set of three-dimensional magneto-hydrodynamics simulations with conditions informed by the parameters of the OMEGA laser facility. We find that jets with a higher degree of helicity may be generated under realistic experimental conditions when compared to a uniform jet. Synthetic x-ray and Thomson scattering diagnostics computed from simulated data demonstrate that the helical jet provides a unique fingerprint in both its morphology and plasma parameters. This laboratory helical jet platform may allow for controlled experimental study of the dynamics of helical plasma structures and, through interaction with other jets or targets, can allow for studies of shear-driven turbulence and mixing relevant to interactions between astrophysical jets and ambient clouds or crosswind.

\end{abstract}

\maketitle
    
\section{Introduction}
\label{section:intro}
Plasma jets are important physical structures that are formed through the ejection of plasma from a source, which are commonly found in astrophysics in young stellar objects (YSOs) containing Herbig-Haro (HH) objects or active galactic nuclei \cite{livioAstrophysicalJetsPhenomenological1999, ciardiCurvedHerbigHaroJets2008a, garciaJetsYoungStars2010, frankJetsOutflowsStar2014}. These astrophysical jets are frequently studied through observation, simulation, and laboratory experiments, investigating questions relating to their formation, collimation and energy dissipation \cite{blackmanPersistentMysteriesJet2022, tzeferacosEffectsEntropyGeneration2013}. Laboratory experiments have been conducted that successfully represent astrophysical jets in a high-energy-density (HED) regime through lasers \cite{gaoMegaGaussPlasmaJet2019} and pulsed-power \cite{blackmanPersistentMysteriesJet2022, stehleScalingStellarJets2009, ciardiLaboratoryStudiesAstrophysical2010}.

Helical jets, defined in this work as jets that have a helical flow velocity or structure, have been confirmed to exist in YSOs \cite{mertensKinematicsJet872016, pudritzRoleMagneticFields2019} and are believed to increase the stability of the jet, especially as it propagates through an ambient medium \cite{ciardiCurvedHerbigHaroJets2008a}. These jets are also important in understanding the angular momentum problem of star formation \cite{woitasJetRotationLaunching2005} by way of the jet carrying away and transporting angular momentum \cite{chrysostomouInvestigatingTransportAngular2008}.  An accurate description of the energy budget of these jets is important to understand such transport, as well as radiative cooling of the jet \cite{tzeferacosEffectsEntropyGeneration2013}. 

Helical, but non-rotating jets, are also of interest in astrophysical systems, namely the jet of SS433 \cite{bowler50SS4332018, monceau-barouxSS433JetSubparsec2015, fabrikaJetsSupercriticalAccretion2006}. The SS433 system consists of a quasar that is precessing while emitting a jet of baryonic matter. This precession causes the jet to have a helical form as it flows outwards into ambient gas. Thus, a helical jet with little rotation is needed in astrophysical laboratory to represent SS433.

Beyond simple helical morphology, magnetic fields in astrophysical plasmas also organize into multiple helical strands that braid around one another, forming complex multi-stranded helical flux rope structures. Such braided configurations have been observed in solar loops \cite{cirtain2013energy}, the Double Helix Nebula \cite{morris2006magnetic} and the M87 jet \cite{pasetto2021reading}. 
Recent theoretical and numerical studies have shown that braided magnetic fields can form stable magnetohydrodynamic equilibria in the form of magnetic double helices, providing a first-principles description of these ubiquitous structures and suggesting that they represent a fundamental magnetic topology in laboratory and astrophysical plasmas \cite{zhang2025magnetic,pawlak2026line}.
Laboratory experiments have reproduced similar braided structures \cite{zhai2015experimental,lavine2021observations}, revealing nontrivial strand–strand interactions \cite{sun2010flux}, intermittent magnetic reconnection \cite{gekelman2016pulsating}, and localized energy release \cite{zhang2023generation}.

\begin{figure}
    \centering
    \includegraphics[]{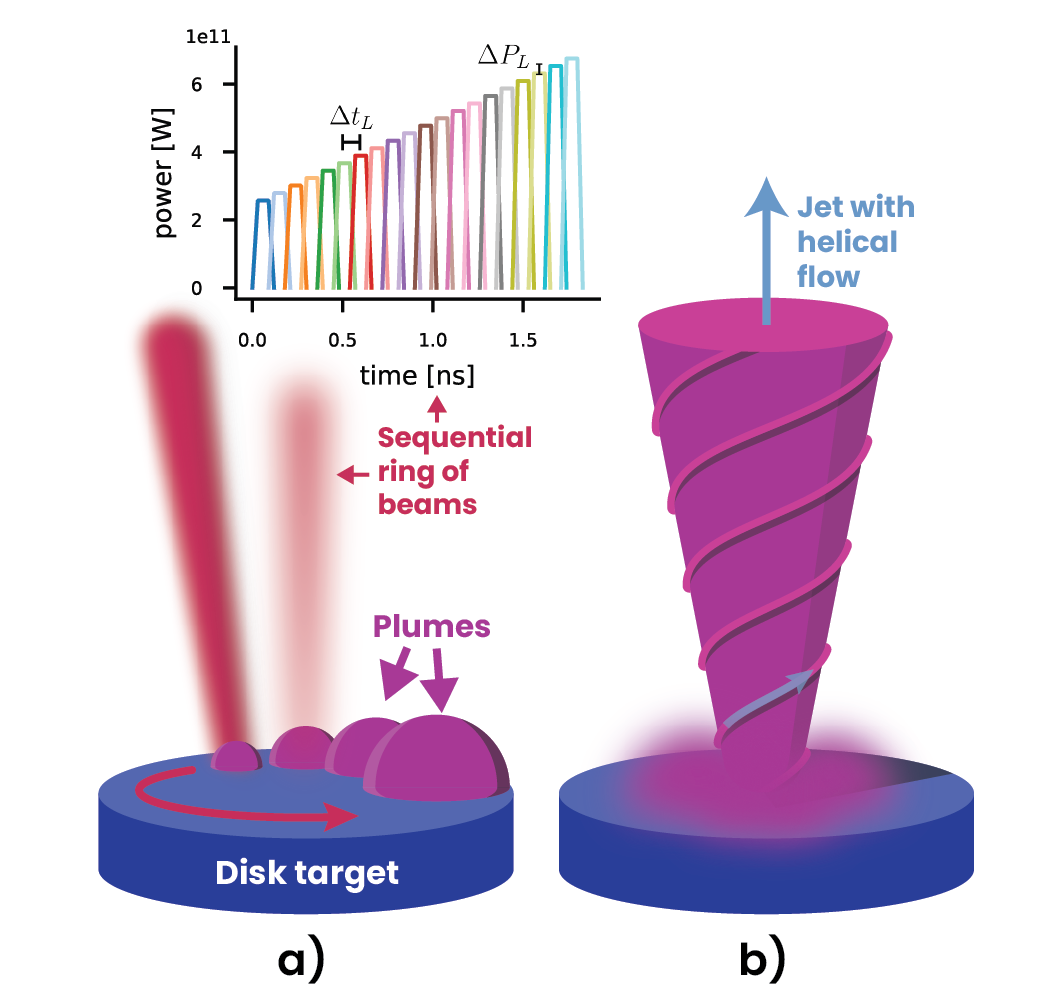}
    \caption{Proposed experimental geometry for the generation of helical jets. A series of laser pulses is incident on a target with varying energy $\Delta P_L$ and time of incidence $\Delta t_L$, depicted in the inset plot, where each color represents a separate beam whose spatial locations are evenly distributed across the ring. This temporal asymmetry imparts an azimuthal velocity, creating a jet with a helical flow on the right.}
    \label{fig:setup}
\end{figure}

Observationally studying any type of astrophysical jet is difficult because they evolve over extremely large temporal and spatial scales, making measurements of the jet dynamics difficult. This is especially true for helical jets whereby the viewing angles inherent to observation make the red and blue shifts unclear, or the helical flow velocity is small, such that it is difficult to determine if the measured shifts are due to helical flows or other unrelated effects \cite{woitasJetRotationLaunching2005}. Thus, having an experimental platform to study a spatially and temporally scaled-down version of helical jets would provide significant utility.

A promising method for generating and understanding the physics of the jet formation and propagation are laser-driven jets. Collimated jets have been driven by ablating a solid target in a ring pattern using several lasers per target \cite{gaoMegaGaussPlasmaJet2019, fuIncreaseDensityTemperature2013, luNumericalSimulationMagnetized2019}, as well as with conical targets \cite{liScaledLaboratoryExperiments2016, gregoryAstrophysicalJetExperiments2008}. The previously studied uniform jets are supersonic with typical temperatures of $T_i>1$ keV, and flow velocities of $v_{\text{jet}}>1000$ km/s. This work will extend the ring-of-beams setup to create a helical jet.

Helical jets have been explored in the laboratory environment in the low density and temperature regime via a plasma gun \cite{lavineHelicalShearFlowStabilization2019}, which proved useful to study the stability that the helical structure has to offer. These plasma gun jets are not able to reach the mach numbers needed to represent YSO jets. Helical jets have also been proven possible in the high energy-density regime with the Z-machine where typical parameters include temperatures of 20 eV, mach numbers of 20 and a low magnetic $\beta=P/P_B$ \cite{amplefordSupersonicRadiativelyCooled2008, cvejicSelfGeneratedPlasmaRotation2022}, whose helical structure is achieved by twisting the Z-pinch's wire arrays. Laser-produced helical jets probe a distinct parameter regime where the electron temperatures are higher than that of the Z-pinch and the kinetic pressure is dominant over the magnetic pressure \cite{gaoMegaGaussPlasmaJet2019}. Typical YSO jets have a wide range of magnetic $\beta$ ($0.1 - 100$), as well as temperatures ($0.1 - 100$ eV) \cite{garciaJetsYoungStars2010} such that the jet discussed in this present work represents the YSO regime with higher kinetic energy as mentioned in section \ref{section:hydro-explanation}. Additionally, laser-produced jets are ideal for experiments because of the open geometry \cite{stehleScalingStellarJets2009}, allowing for more diagnostic flexibility. Lasers also offer flexibility in plasma parameters, as changing the beam power, timing, and the ring-of-beams radius results in a range of densities, temperatures and velocities.

Leveraging the experimentally demonstrated laser-driven uniform jets, in this manuscript we explore a platform for generating helical jets. Such a configuration can be realistically driven by the lasers at the OMEGA laser facility at the University of Rochester Laboratory for Laser Energetics \cite{boehlyInitialPerformanceResults1997}. We demonstrate how such a jet may be diagnosed by generating synthetic x-ray images and spatially averaged velocity profiles as would be obtained from Thomson scattering. The results of this work will allow for future experimental studies of helical jets where they may be studied in isolation or interacted with other jets or targets to provide insight into instability formation and mixing.

\section{Experiment Setup}
\label{sec:experiment-setup}
Prior work on laser-driven jets has successfully demonstrated the propagation of magnetized jets, from which we will build here \cite{fuIncreaseDensityTemperature2013, luNumericalSimulationMagnetized2019, gaoMegaGaussPlasmaJet2019}. In previous experimental and simulation work, uniform jets were created by irradiating a plastic target with a ring of beams that merge at the center of the target to form a jet, as explained in section \ref{section:hydro-explanation}.
    
Fig. \ref{fig:setup} demonstrates the concept for the simulations performed to create a helical jet. In this work, 3-dimensional magneto-hydrodynamic (MHD) simulations are performed with 20 beams distributed in the shape of a ring onto a plastic (CH) disk of radius 12 mm and 5 mm thickness and arrive on-target sequentially. The radius of the ring is chosen to be 800 microns, consistent with previous simulation and experimental work on OMEGA \cite{gaoMegaGaussPlasmaJet2019, luNumericalSimulationMagnetized2019}. Each beam is turned on and then off sequentially over time, as in Fig. \ref{fig:setup}(a). Each sequential pulse is given more power to prevent the $n\text{th}$ beam from dominating over the $n\text{th}+1$ beam, where $n$ is a single beam in the sequence of beams. To adhere to the experimental constraints of OMEGA, the max power used for an individual beam is $0.7 \text{ TW}$, totaling  $\sim 0.85 \text{ TW}$ when summing all beams.

The laser configuration for the helical jet can be parametrized as follows. The laser pulses arrive on target in a ring at the same angle of 30$^{\circ}$ to the target-normal. The lasers have a full-width-half-maximum (FWHM) of 250 $\mu$m, providing a small spacing between each of the laser spots. The pulses have a delay between pulses $\Delta t_L$, which is depicted in the inset laser profile of Fig. \ref{fig:setup}(a). The change in pulse energy $\Delta P_L$ between each beam may also be varied. For this work, we only consider $\Delta t_L = 0.09$ ns and a linear increase in the laser energy of $\Delta P_L = 2.2 \times 10^{11}$ W between subsequent pulses, ending with $6.75 \times 10^{11}$ W at $t = 1.8$ ns, totaling to $\sim 1$ kJ of laser energy when summing over all beams. Each individual beam pulse is 0.12 ns and ranges in power from $23$ kJ for the first beam to $61$ kJ for the last beam. The uniform jet ($\Delta t_L = 0$ ns, $\Delta P_L = 0$ W), to which the helical jet will be compared, has a $1$ ns pulse consistent with previous studies, but at a lowered total energy of $1$ kJ to match the helical jet. The several laser parameters allow for control over the plasma parameters of the jet, including the velocity, density and helicity.

\section{Simulation Setup}
\label{section:flash-setup}

To study the generation of helical jets formed under various laser configurations, the magnetohydrodynamics code FLASH \cite{fryxellFLASHAdaptiveMesh2000} is employed. This multiphysics code is commonly used for long-pulse laser interactions with nanosecond timescales due to the collisionality of these systems and the inclusion of laser energy deposition. We note that the systems considered here are collisional and therefore the use of FLASH is valid. Laser deposition in FLASH is modeled by ray-tracing coupled with collisional inverse Bremsstrahlung heating, which is the dominant heating mechanism for this laser intensity regime. Magnetic fields are included such that when this laser energy is deposited, Biermann Battery fields are formed, as discussed in section \ref{section:magnetic-fields}. The background within the simulation is filled with low-density helium ($2\times10^{-7}$ g/cm$^3$) to avoid having an unsolvable Riemann problem while also mimicking the vacuum conditions present in experiment. Although necessary, this sharp density and temperature gradient causes the jet to shock the helium (the jet is flowing faster than the sound speed of the helium). To avoid unnatural Biermann fields at this shock front, the magnetic field is not computed at these shock regions.

To represent the CH target and helium material, IONMIX equation of state files are used. These contain tabulated densities and temperatures with corresponding parameters, like pressures and average ionization, that are then used to interpolate the same parameters for a given density and temperature in the FLASH simulation. As the plastic jet is fully ionized, it acts close to an ideal gas. These equations of state provide closure to the included MHD equations, allowing for a full simulation with MHD and laser energy deposition.

\section{Results}
\label{section:results}
\subsection{Hydrodynamics}
\label{section:hydro-explanation}

\begin{figure}
    \centering
    \includegraphics[]{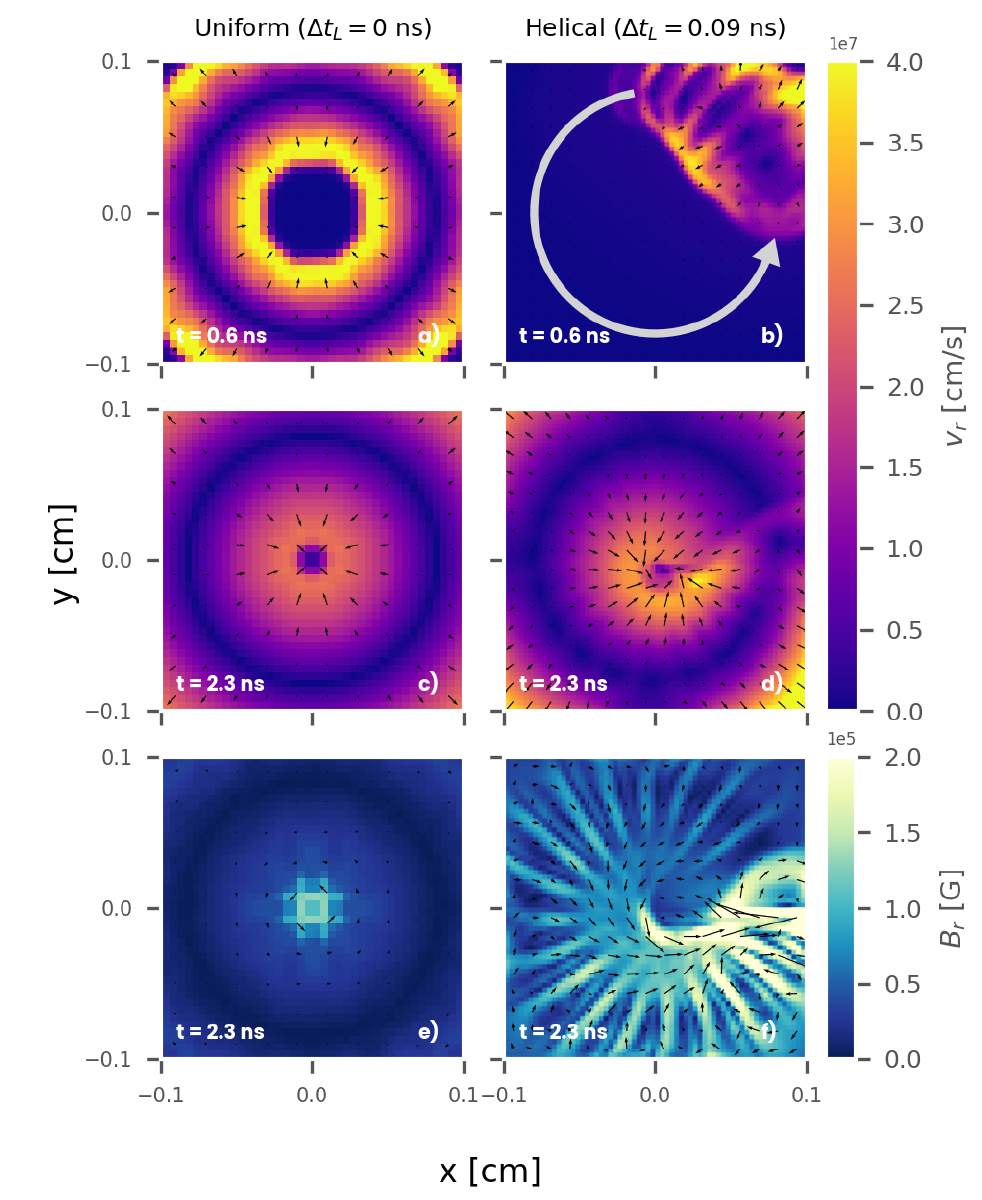}
    \caption{Top-down view $0.05$ cm above the target of the radial velocity $v_r = \sqrt{v_x^2 + v_y^2}$ at $t = 0.6$ ns, as well as $v_r$ and $B_r = \sqrt{B_x^2 + B_y^2}$ at $t = 2.3$ ns, for each type of jet with the same total amount of laser power. \textbf{(a)} The uniform jet uses beams that are all incident on the target at the same time, causing the azimuthal velocities to effectively cancel and \textbf{(c)} leaving only inward and outward radial components. \textbf{(b)} In contrast, the helical jet is produced through sequential beams, such that the azimuthal velocities do not cancel, \textbf{(d)} producing a net azimuthal velocity while the beams come back around to the start. The azimuthal magnetic fields in \textbf{(e)} emerge from the cancellation of the plumes' radial Biermann fields, which are not fully canceled in \textbf{(f)}.}
    \label{fig:helical-vs-uniform}
\end{figure}

To illuminate the difference between the uniform ($\Delta t_L = 0 $ ns) and helical jets ($\Delta t_L = 0.09$ ns), Fig. \ref{fig:helical-vs-uniform} plots slices of each jet near (0.05 cm above) the target surface, displaying a crucial point in each of the jets' formation. In Fig. \ref{fig:helical-vs-uniform}(a), the simultaneous arrival of beams generate plumes that eventually merge at the center. The plumes are nearly identical and close together, causing azimuthal velocities to effectively cancel and leave two rings with only radial velocity in the plane: one that expands outward out of the system and one that travels inward. The latter is what will eventually collide at the center of the target to create the uniform jet. At the central collision point of the plumes, radial velocities cancel, while target-normal velocities are unaffected, driving a collimated jet that rapidly propagates away from the target.

In the case of the sequential beams, the azimuthal velocity of each plume is less than the plume before it. A net non-zero azimuthal velocity is thus generated around the ring of plumes that is carried into the central region and out from the target surface by the plumes' axial velocity, creating a helical jet. In Fig. \ref{fig:helical-vs-uniform}(d), the quiver arrows show a clear net angular momentum created by the asymmetric plasma plumes in the case of the helical jet and only a radial velocity for the uniform irradiation in Fig. \ref{fig:helical-vs-uniform}(b). As the sequential beams stop after their first ring is completed, this angular momentum is not continuously injected.

\begin{figure*}
    \centering
    \includegraphics[]{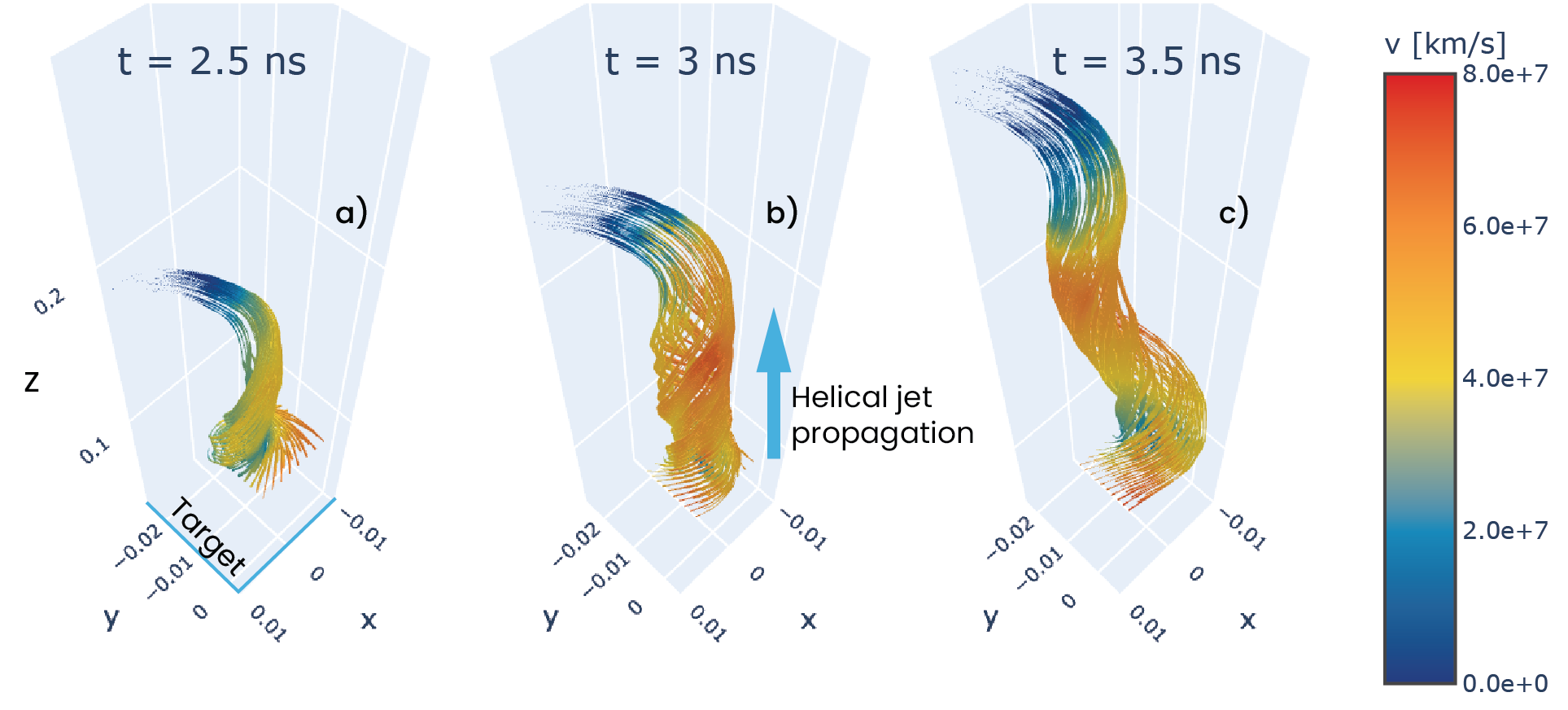}
    \caption{3D velocity streamlines of a central region of the jet that follow a helical path and evolve over time as the jet propagates upwards, well after the last beam has turned off at $t = 1.8$ ns. The color of the streamlines indicates the magnitude of the flow velocity. (a) The helical structure starts to take shape at $t = 2.5$ ns, (b) continues to evolve and (c) reaches a full period.}
    \label{fig:streamlines}
\end{figure*}

This azimuthal velocity is then brought up by the axial flow of the original flow, creating a helical flow velocity, whose streamlines are shown in Fig. \ref{fig:streamlines}. In Fig. \ref{fig:streamlines}(a), the jet is off-centered due to the increasing beam arrival times and beam powers, the reasons for which are explained in section \ref{sec:experiment-setup}, causing the total energy for each beam to be unequal and pushing the center of the jet away from center of the $x-y$ plane. The jet is is initially $0.01$ cm wide and widens over time due the high pressure gradient between the jet and the vacuum in Fig. \ref{fig:streamlines}(b) and (c), as it propagates into vacuum to $0.3$ cm above the target.

\begin{figure}
    \centering
    \includegraphics[]{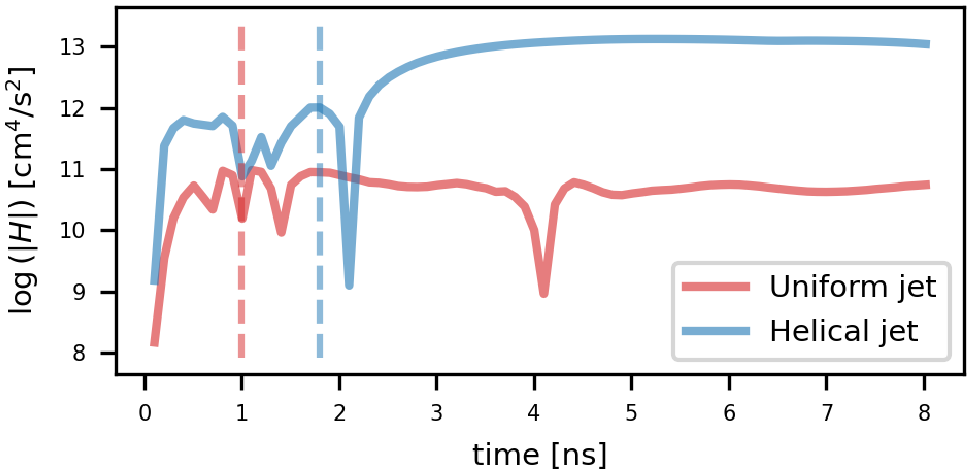}
    \caption{The helicity of the helical jet is magnitudes higher than that of the uniform jet, made clear by looking at the log of $|H|$ over time. The vertical dotted lines indicate when the beams turn off for each respective simulation. When the final beam turns off, the helical jet retains a consistent helicity, while the uniform jet's helicity is inconsistent. The helicity of the uniform jet is effectively zero as the jet is symmetric, but due to the asymmetry caused by a finite number of computational cells, the helicity does not fully cancel over the volume integral.}
    \label{fig:spacetime}
\end{figure}

The resulting jet sustains a helical structure throughout it's evolution, as in Fig. \ref{fig:streamlines}, with a consistently higher kinetic helicity 
\begin{equation}
    H = \int{v \cdot \omega\text{d}V},
\end{equation}
where $v$ is the flow velocity and $\omega = \nabla \times v$ is the vorticity, compared to the uniform jet. For a structure to have helical flow, vorticity must follow the direction of velocity, such that $|v \cdot \omega| > 0$. Helicity is an indication of helical structure because $\omega$  gives a measure of the vortical flow, allowing $v \cdot \omega$ to provide an idea how much this vortical flow is aligned with the bulk axial flow. If there is a bulk axial flow (the jet), along with a vortical flow, then a helical flow is created. Helicity, being the volume integral of $v \cdot \omega$, is telling of the alignment between the flow velocity and vorticity and therefore is a measure of how helical the structure is. In Fig. \ref{fig:spacetime}, the helicity in the helical jet case is not only magnitudes higher than the uniform case, but also sustains its helicity throughout the simulation, indicating a strong helical structure and flow. To give a reference value for Fig. \ref{fig:spacetime}, we may calculate a characteristic helicity $H_{\text{ref}}\sim \frac{C_s^2}{R_{\text{jet}}}V = \pi C_s^2 R_{\text{jet}}l_{\text{jet}} \sim 10^{12}$ cm$^4$/s$^2$, where $c_s = \sqrt{\gamma P/\rho} \sim 10^7 $ cm/s is the sound speed of the jet, $V$ is the cylindrical volume of radius $R_{\text{jet}}$, $R_{\text{jet}}\sim 0.01$ cm is the radius of the jet, and $l_{\text{jet}}\sim0.35$ cm is the height of the jet, for the helical jet case. The value of $H_{\text{ref}}$ is lower by a magnitude compared to the maximum in Fig. \ref{fig:spacetime}, due to the fact that Fig. \ref{fig:spacetime} integrates over the simulation domain rather than the smaller volume of the jet, meaning that $H_{\text{ref}}$ and the helical jet case of Fig. \ref{fig:spacetime} are in agreement. Altering the laser parameters mentioned in section \ref{sec:experiment-setup} would increase or decrease the velocity, providing control over the helicity of the jet.

The platform also offers versatility in torque. If there is a helical flow, then there must be non-zero torque, $\mathbf{\tau} = r\mathbf{F}$, where $r$ is the lever arm radius and $\mathbf{F}$ is the force. Each individual plume has zero torque themselves, but when considering all of the plumes at the global scale, the net torque is non-zero, where the force is the pressure gradients between the plasma plumes and the lever arm radius is the ring-of-beams radius. In other words, integrating the system over time would output a net zero torque, but when looking at individual time slices, there is a net torque. Integrating the plumes' torque of the helical jet over time is the same as having the beams arrive simultaneously, as is true in the uniform jet case where there is zero net torque. So, the timing between each of the beams is vital to the torque put upon the jet and serves as an additional tuning parameter. Additionally, increasing or decreasing the radius will also alter the amount of torque present in the system, along with the pressure gradients between plumes if not spaced the same.

The helical structure propagates at supersonic speeds up to internal mach numbers $M = v_{\text{flow}} / c_s$ of 5, competing with its uniform counterpart. For this particular beam configuration, the electron temperature of the jet reaches 200 eV as measured by FLASH (less than previous studies due to less total energy) and high magnetic $\beta$, providing a regime not available to other laboratory jets, including pulsed power.

\subsection{Magnetic fields}
\label{section:magnetic-fields}
In laser-target interactions, a magnetic field is created via the Biermann Battery effect \cite{Biermann_PR_1951} and is represented as the last term here in generalized Ohm's law as
\begin{equation}
    \frac{\partial B}{\partial t} = -\nabla \times \left(-\mathbf{v} \times \mathbf{B}  + \eta \mathbf{j}\right) -\frac{\nabla n_e \times \nabla T_e}{en_e},
\end{equation}
where non-collinear gradients in electron temperature $T_e$ and density $n_e$ lead to the generation of a magnetic field $B$. This misalignment is inherent to the laser-target setup and creates clockwise azimuthal fields around each plume via a large downward $\nabla n_e$, created by the rocket effect, and $\nabla T_e$ along the target surface created by the gaussian laser pulse shape, which both ultimately lead to a clockwise azimuthal field around each plume.

The azimuthal fields experience something similar to the velocity vectors at startup, as discussed in section \ref{section:hydro-explanation}. Normally in the formation of a uniform jet, the magnetic field radial components at the intersection of two plumes cancel, leaving only an inner and outer ring that run azimuthally clockwise and counter-clockwise, respectively, when looking down onto the target surface as in Fig. \ref{fig:helical-vs-uniform}(e). However, the radial magnetic field components of the helical jet do not fully cancel, which creates a radial field pointing inwards in Fig. \ref{fig:helical-vs-uniform}(f). 

The field from the last plume created is hot as laser energy has recently been deposited while the first plume has cooled down by the last laser pulse. This last hot plume and cold first plume interact, which creates a large temperature gradient and thus a large Biermann field forms on the right-hand side of Fig. \ref{fig:helical-vs-uniform}(f) affecting the organization of the helical structure, such that purely hydrodynamic simulations exhibit a more consistent helical structure than with MHD simulations.

Nernst effect advects magnetic fields down temperature gradients \cite{walshExtendedmagnetohydrodynamicsUnderdensePlasmas2020} but is not included in these simulations. 
It is important for the evolution of the magnetic field but not for the global structure of the jet \cite{luNumericalSimulationMagnetized2019} because of the high-beta regime of the jet. At start-up in the helical jet case, the magnetic field would be advected further down the temperature gradient created by the last and first plume. Once the jet is formed, temperature gradients are minimal and would only have small-scale effects on the magnetic field.

\subsection{Synthetic diagnostics}
Now that we have demonstrated how helical jets can be formed, we will demonstrate how they can be measured. To do this we will consider two commonly used diagnostics: x-ray pinhole imaging and Thomson scattering.

\subsubsection{Imaging of x-ray self-emission}
\begin{figure}
    \centering
    \includegraphics[]{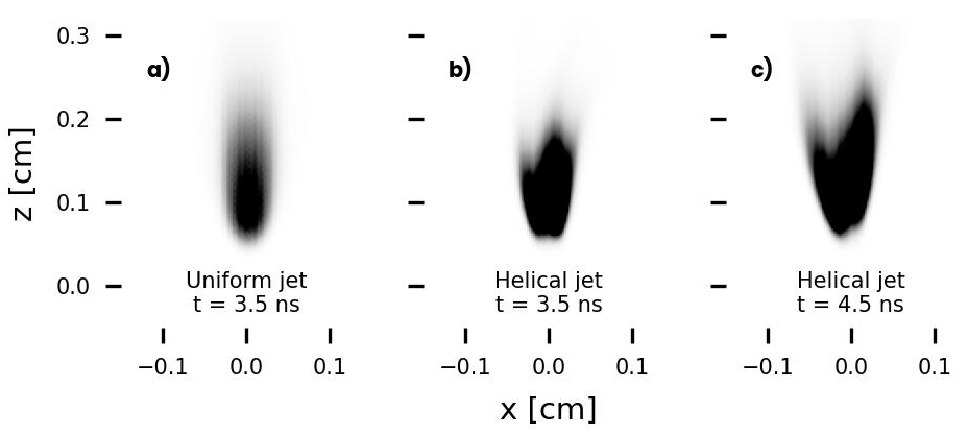}
    \caption{Synthetic x-ray self-emission of the uniform jet at $t = 3.5$ ns and helical jet at both $t = 3.5$ ns and $t =4.5$ ns. \textbf{(a)} Uniform jet with 1 kJ of energy whose morphology is similar to previous studies of uniform jets \cite{gaoMegaGaussPlasmaJet2019}. \textbf{(b)} Predicted output of x-ray self-emission for the helical jet at $3.5$ ns and \textbf{(c)} $4.5$ ns.}
    \label{fig:synth_xrays}
\end{figure}

The x-ray emission from the plasma itself is modeled as bremsstrahlung radiation using $T_e$, $n_e$ and $n_i$ values from FLASH. The emission from each FLASH voxel is projected onto each pixel of a detector with an emission strength proportional to $r_{ij}^{-2}$ where $r_{ij}$ is the distance from the FLASH voxel $i$ to the detector pixel $j$. In between the detector and the simulation domain is a pinhole of radius $0.0025$ cm that is necessary to focus the image of the jet onto the detector. Since the target is composed of low-$Z$ polystyrene, the jet itself is fully ionized and experiences very little line radiation such that emission can be readily modeled with bremsstrahlung radiation

\begin{equation}
    j_{ff} = 2 \times 10^{-32}Z^2n_en_i\exp{\left(-\varepsilon/T_e\right)}T_e^{-1/2},
\end{equation}

where $\varepsilon$ is the photon energy.
The results of this synthetic diagnostic are shown in Fig. \ref{fig:synth_xrays}. Fig. \ref{fig:synth_xrays}(a) is synthetic x-rays of the uniform jet, displaying similar morphology to previous studies \cite{gaoMegaGaussPlasmaJet2019}, giving confidence towards the synthetic output.

To confirm the helicity of the jet in an experiment, it has to look and behave like a helical jet. In Fig. \ref{fig:synth_xrays}(b) and (c), the synthetic x-rays are computed for the helical jet, where strands from the tracing out of a cone-shape are seen at the top of the jet, providing a unique fingerprint for an experimental helical jet. These strands are reminiscent of the strands seen in interferometry measurements in previous helical jet experiments on the z-machine \cite{amplefordSupersonicRadiativelyCooled2008} and astrophysical simulations of SS433 \cite{monceau-barouxSS433JetSubparsec2015}, as they are caused by the jet tracing out a cone shape, corroborating what the jet should \textit{look} like. To image and measure this shape, the resolution of the detector in the experiment needs to be high enough to resolve the strands.
The detection of x-ray self-emission provides spatial and temporal understanding of the jet critical to the placements of Thomson-Scattering diagnostics. 

\subsubsection{Thomson scattering}

While x-rays may be used to provide a qualitative picture of the jet structure, additional diagnostics should be used to to provide a concrete measurement of how the jet dynamically \textit{behaves} and to confirm that it is helical through density, temperature and velocity measurements. For this purpose, streaked collective Thomson scattering is an ideal diagnostic because it can provide time resolved measurements of all the previously mentioned properties \cite{Froula_FST_2012}. By scattering a laser pulse from a plasma and collecting the scattered light into two temporally streaked spectrometers, the evolution of the narrow-band spectral features resulting from the interaction with ion acoustic waves (IAW) and the more broadband spectral features resulting from electron plasma waves (EPW) may be diagnosed. The laser spot size and collection optics define the scattering volume and the angle between the probe and scattered light defines the k-vector of the waves being measured, and therefore the propagation direction of the flows being diagnosed. Using the known functional form for the EPW and IAW, fits may be performed to obtain the electron density, electron temperature, ionization state, electron drift velocity, bulk plasma flow velocity, and the ion temperature. This has been used in many HEDP experiments to provide robust measurements of plasma parameters \cite{Schaeffer_PRL_2019, Fiuza_NatPhys_2020,Zhang_NatPhys_2023} and it is a standard diagnostic at the OMEGA laser facility. For the purpose of the proposed helical jet platform, collective Thomson scattering may be used to confirm that it is helical. 

\begin{figure}
    \centering
    \includegraphics[]{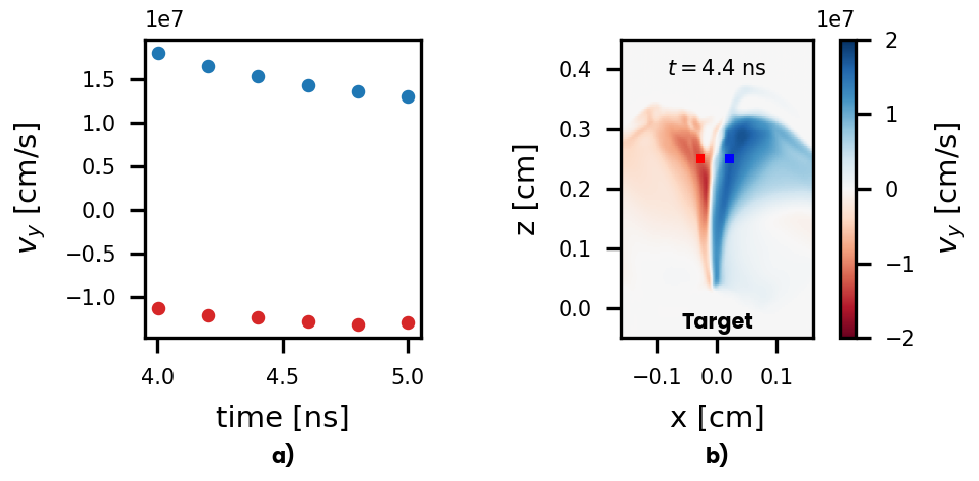}
    \caption{Prediction of Thomson Scattering data using FLASH data. Each side of the helical jet is probed to measure the velocity going into and out of the page. \textbf{(a)} Spatially integrated measurements over 50 microns, indicated by the red and blue boxes in (b), at different times. The blue (out of the page) and red (into the page) velocities are measurements on the left and right side, respectively, of the helical jet in (b). \textbf{(b)} Slice at $y=0$ cm at 4.4 ns. Each side of the helical jet is clearly going in the opposite direction to the other side of the jet.}
    \label{fig:sythetic-thomson}
\end{figure}

To demonstrate that the jet has a helical flow, collective Thomson scattering may be employed whereby flow velocity can be diagnosed through red- and blue-shifts in the IAW feature. In Fig. \ref{fig:sythetic-thomson}a, the FLASH in- and out-of-plane velocity ($v_y$ in this case) is averaged in a 50-micron cubed volume over time to approximate the scattering volume from a Thomson scattering measurement, resulting in a clear separation in positive (out of the page) and negative (into the page) flows, both with magnitudes of $100$ km/s. These out-of- and in-plane velocities are seen clearly in Fig \ref{fig:sythetic-thomson}b, where each half of the jet is flowing opposite to the other. Experimentally, this could be done in two shots where the  placement of the probe laser will be informed by the XRFC images, as they will help in identifying where the velocity should be positive or negative.

\section{Discussion and Conclusions}
\label{section:discussion}

A helical jet formed from laser irradiation offers flexibility in studying both astrophysical phenomena and fundamental HED plasmas physics. 
This laser-produced helical jet has a helical flow, but helical magnetic fields are also of interest within astrophysical jets \cite{pasetto2021reading}. Given that there are both azimuthal and axial magnetic fields in the laser-produced uniform jet \cite{luNumericalSimulationMagnetized2019}, one would expect a helical field to form naturally within a laser-produced uniform jet, but the azimuthal field comes up short as it exists most strongly close to the target and is otherwise advected weakly throughout the jet \cite{GenerationAstrophysicsrelevantHelical}. With the laboratory formed helical jet, it is possible to create a helical magnetic field by applying an axial magnetic field that is swept up by the helical flow, making it beneficial for modeling helical magnetic fields in astrophysical jets.

Commonly in astrophysics, a jet propagates into an ambient medium or crosswind \cite{ciardiCurvedHerbigHaroJets2008a}, shearing and causing shocks and instabilities to grow. This case is relevant to HH objects and especially the SS433 system where the system's highly supersonic jet punches into the surrounding nebula \cite{bowler50SS4332018}. Representing this scenario with this laser-produced helical jet is possible by adding an ambient gas through which the jet would propagate.


The helical jet platform can be potentially applied in studies of turbulence \cite{tzeferacosLaboratoryEvidenceDynamo2018} and mixing \cite{dimotakisTURBULENTMIXING2005a, arnettRoleMixingAstrophysics2000} relevant to both astrophysics, such as in supernovae \cite{meakinTurbulentConvectionStellar2007} and fundamental studies of mixing in HED plasmas \cite{dossInstabilityMixingTransition2013, benderSimulationFlowPhysics2021, zhouInstabilitiesMixingInertial2025}.
The path to turbulence is made clear through the growth of instabilities that create smaller and smaller scales, eventually leading to mixing. The helical jet is a promising way of injecting energy towards instability growth, specifically through shear of velocities. When the helical jet penetrates ambient gas, for example, the azimuthal and axial flow will shear against the smaller velocity of the gas. 
A large gradient in velocity is also possible through the collision of two helical jets that have azimuthal flows opposite to each other (opposite helicities). These gradients in velocity at every point of the collision interface could potentially lead to Kelvin-Helmholtz instabilities, generating turbulence and mixing of target materials. The increased helicity within the jet collision would also contribute towards the sustainment of the magnetic fields against resistive decay, contributing towards the magnetic dynamo.

Laser-driven plasma flows and jets have been used in various configurations to study turbulence and the magnetic dynamo on OMEGA \cite{tzeferacosLaboratoryEvidenceDynamo2018}, where sub-sonic colliding flows shear to create turbulence and a dynamo. Colliding streams of plasma have been explored further \cite{huntingtonMagneticFieldProduction2017, fiuzaElectronAccelerationLaboratoryproduced2020}, making the adaptation of the single helical jet platform to a colliding helical jet platform feasible.
This platform offers flexibility in how the setup can be varied, including several beam parameters and target design, in addition to the richness in physics relevant to astrophysics and HED studies.

\section*{Acknowledgments}

This work has been supported by Princeton University/Princeton Plasma Physics Laboratory under contract number DE-AC02-09CH11466 with the U.S. Department of Energy through a DOE Early Career Research Program grant DOE-FOA-0003176. The United States Government retains a non-exclusive, paid-up, irrevocable, world-wide license to publish or reproduce the published form of this manuscript, or allow others to do so, for United States Government purpose(s). B.K.R. was supported by the US Department of Energy High-Energy-Density Laboratory Plasma Science program under Grant No. DE-SC0020103. Y.Z. was supported by the NASA Living with a Star Jack Eddy Postdoctoral Fellowship Program, administered by UCAR's Cooperative Programs for the Advancement of Earth System Science (CPAESS) under award $\#$80NSSC22M0097. 

This report was prepared as an account of work sponsored by an agency of the United States Government. Neither the United States Government nor any agency thereof, nor any of their employees, makes any warranty, express or implied, or assumes any legal liability or responsibility for the accuracy, completeness, or usefulness of any information, apparatus, product, or process disclosed, or represents that its use would not infringe privately owned rights. Reference herein to any specific commercial product, process, or service by trade name, trademark, manufacturer, or otherwise does not necessarily constitute or imply its endorsement, recommendation, or favoring by the United States Government or any agency thereof. The views and opinions of authors expressed herein do not necessarily state or reflect those of the United States Government or any agency thereof.

\section{Author Declarations}
\subsection{Conflict of Interest}
The authors have no conflicts to disclose.

\section*{Data Availability}

The data that support the findings of this study are available from the corresponding author upon reasonable request.

\bibliography{compressed_bib,Helical-jet, LRGF-fellowship, HEDP_all}

\end{document}